\documentclass[titlepage,12pt]{article}
\usepackage{amssymb,epsfig,pslatex,psfrag}
\usepackage{hyperref}
\usepackage[english]{babel}
\usepackage[sans]{dsfont}
\usepackage[latin1]{inputenc}
\usepackage[T1]{fontenc}
\makeatletter
\def\@sect#1#2#3#4#5#6[#7]#8{\ifnum #2>\c@secnumdepth
  \def\@svsec{}\else 
  \refstepcounter{#1}\edef\@svsec{\csname the#1\endcsname.\hskip0.5em}\fi
  \@tempskipa #5\relax
  \ifdim \@tempskipa>\z@
    \begingroup 
      #6\relax
      \@hangfrom{\hskip #3\relax\@svsec}{\interlinepenalty \@M #8\par}%
    \endgroup
    \csname #1mark\endcsname{#7}\addcontentsline
      {toc}{#1}{\ifnum #2>\c@secnumdepth \else
        \protect\numberline{\csname the#1\endcsname}\fi #7}%
  \else
    \def\@svsechd{#6\hskip #3\@svsec #8\csname #1mark\endcsname
      {#7}\addcontentsline{toc}{#1}{\ifnum #2>\c@secnumdepth \else
        \protect\numberline{\csname the#1\endcsname}\fi #7}}%
  \fi \@xsect{#5}}

\renewcommand{\>}{\rangle}

\newcommand{\one}{1\!\mbox{l}}
\newcommand{\bfsig}{{\mbox{$\sigma$}}}

\def\one{\mathds{1}}
\begin{document}
\begin{titlepage}
  \begin{flushright}
    April 2003 \\
    DESY 03-56\\
    PITHA 03/03\\
    TTP03-14\\    
  \end{flushright}

\vspace*{0.4cm}

\renewcommand{\thefootnote}{\footnotesize\fnsymbol{footnote}}
\begin{center}
  {\LARGE\bf Spin properties of top quark pairs produced
    at hadron colliders\footnote{Invited talk given at the 
      Cracow epiphany conference on heavy flavours,
      3 - 6 January 2003, Cracow, Poland.}} \\
\vspace*{2cm}

{\bf W. Bernreuther  $^{a}$, 
  A. Brandenburg $^{b,}$\footnote{Speaker at the conference.}, 
  Z.G. Si $^{c}$, P. Uwer $^{d}$
  }
\vspace*{1cm}

$^a$  Institut für Theoretische Physik, RWTH Aachen, 52056 Aachen, Germany \\
$^b$  DESY-Theorie, D-22603 Hamburg, Germany\\
$^c$  Department of Physics, Shandong University, Jinan Shandong
250100, China\\
$^d$  Institut für Theoretische Teilchenphysik, Universität Karlsruhe, 
76128 Karlsruhe, Germany
\vspace*{3cm}

{\bf Abstract:}\\
\parbox[t]{0.8\textwidth}
{We discuss the spin properties of top quark pairs produced at hadron colliders
at next-to-leading order in the coupling constant $\alpha_s$ of the strong
interaction. Specifically
we present, for some decay channels,  results
for differential angular distributions  that
are sensitive to $t\bar t$  spin correlations.}
\end{center}
\vspace*{2cm}

PACS number(s): 12.38.Bx, 13.88.+e, 14.65.Ha

\end{titlepage}

\renewcommand{\thefootnote}{\arabic{footnote}}
\setcounter{footnote}{0}

\section{Introduction}
The top quark is the heaviest fundamental particle discovered
so far. Its interactions are still relatively unexplored and their 
experimental investigation may lead to exciting new discoveries. 
To mention only a few examples:
due to its large mass, it is not excluded 
that the top quark decays into yet unobserved particles
like charged Higgs bosons or supersymmetric particles 
-- or, if these particles are heavier than the top quark, they may
mediate top quark decay, leading also to new decay modes and/or
branching ratios that differ from the Standard Model (SM) predictions
(for overviews, see, e.g., refs. \cite{Beneke:2000hk,Accomando:1997wt}).
Specifically, the V$-$A structure of the top quark
decay vertex is modified in many extensions of the Standard Model 
\cite{Soni:1992tn,Cho:zb}. 
In the production of top quarks, the resonant production of heavy 
spin-0-particles could lead to interesting signatures 
\cite{Bernreuther:1993hq,Dicus:bm,Bernreuther:1997gs}. 
Its large mass makes the top quark also a good probe of the electroweak 
symmetry breaking mechanism. In this context it will be important 
to check whether the 
Yukawa coupling of the top quark is as predicted by the Standard Model.
Finally, the question whether the discrete symmetry CP is violated
in top quark production and decay  has been the subject of 
numerous investigations (cf. ref. \cite{Atwood:2000tu}
for a review).
The foreseen large data samples at the upgraded Tevatron 
($\sim 10^3-10^4\ t\bar{t}/$year) and at the LHC 
($\sim 10^7\ t\bar{t}/$year) will tell us more about these 
issues in the near future.
Of help in this context will be
another special feature of the top quark, namely that
the spin properties of top quark pairs are predictable
by perturbation theory. This is due to  its large
decay width, $\Gamma_t^{\rm SM}\sim 1.5$ GeV 
$\gg \Lambda_{\rm QCD}$, which serves 
as a cut-off for hadronization effects. 
In particular, the top quark
decays so fast that the information about its polarization is not 
diluted by hadronization but transferred to the decay products.
Thus observables related to the  spins of the $t$ and $\bar{t}$ quarks
can be constructed and used reliably for the
detailed study of the dynamics of top quark
production and decay \cite{Kuhn:1983ix}. 
Apart from searching for non-standard effects, the study of the $t$
and $\bar{t}$
spin properties is interesting even within the framework of the SM: 
They probe the `quasi-free' nature of the top quark, thus allowing us
to study properties of a `bare' quark. Further, a measurement of
spin correlations would provide a lower bound on $|V_{tb}|$ without
assuming the existence of three quark generations 
\cite{Stelzer:1996gc}. 

Here we confine ourselves to top quark pair
production and decay at hadron colliders and investigate  
the top quark polarization and spin correlation  phenomena that
are induced by the strong interaction dynamics. 
We discuss predictions  for the normal polarisation of the
top quarks and $t \bar t$ spin correlations at
next-to-leading order (NLO) in the QCD coupling $\alpha_s$.

\section{Theoretical framework}
Needless to say, for a correct interpretation of upcoming
and future data on top quark production and decay
at the Tevatron and the LHC,   
precise theoretical 
predictions 
within the SM are needed.
We consider here the following reactions:
\begin{equation}
  h_1h_2\to t\bar{t}+X\to
  \left\{\begin{array}{lcc}
      \ell^+\ell '^{-} &+& X\\ 
      \ell^+j_{\bar{t}} &+& X\\ 
      \ell^-j_{t} &+&X\\
      j_t j_{\bar{t}}&+&X
    \end{array} \right.,
  \label{reac1}
\end{equation}
where $h_{1,2}=p,\bar{p}$; $\ell=e,\mu,\tau$, and 
$j_t\ (j_{\bar{t}})$ denotes  
jets originating from hadronic $t$ ($\bar{t}$) decays. 
Experimental analysis of the above processes
requires predictions of the fully differential cross sections.
The calculation of these cross sections at  next-to-leading order
 QCD simplifies 
enormously in the leading pole
approximation (LPA), which amounts to expanding the full
amplitudes for the reactions listed in eq.~(\ref{reac1}) 
around the complex poles of the
top quark propagators. Only the leading pole terms are kept in this 
expansion, i.e.,  one neglects terms of order $\Gamma_t/m_t\approx 1\%$.
Within the LPA, the radiative corrections can be classified into
factorizable and non-factorizable contributions.
Here we consider only the factorizable corrections; for
the non-factorizable contributions see  ref. \cite{Beenakker:1999ya}.
Further we apply the on-shell aproximation for the 
top quark propagators:
\begin{eqnarray}
  \lim_{\Gamma/m\to 0}|\frac{1}{k^2-m^2+im\Gamma}|^2\to
  \frac{\pi}{m\Gamma}\delta(k^2-m^2).
\end{eqnarray}
The necessary ingredients at  NLO QCD within this approximation are
the differential cross sections for the following
parton processes
\begin{eqnarray}
  q\bar{q}\to t\bar{t},\ gg \to t\bar{t} 
  && \mbox{(to order $\alpha_s^3$)},\\ 
  q\bar{q}\to t\bar{t}g,\ gg \to t\bar{t}g \ 
  && \mbox{(to order $\alpha_s^3$)},\\
  g q(\bar{q})\to t\bar{t}q(\bar{q})
  && \mbox{(to order $\alpha_s^3$)},\\
  t\to b\ell\nu,bq\bar{q}'
  && \mbox{(to order $\alpha_s$),}
\end{eqnarray}
where we have to keep the full information on 
the $t$ and $\bar{t}$ spins.

\section{Spin density matrix}
The fully differential
cross section $d\sigma_{\rm fact.}$ for the production and subsequent decay
of top quark pairs at the parton level can be written in terms of 
a spin density matrix $R$ and decay density matrices $\rho,\bar{\rho}$: 
\begin{equation}
d\sigma_{\rm fact.}={\rm Tr}_{t,\bar{t}\ \rm spins}\left(R\ \rho\ \bar{\rho}\right).
\end{equation} 
The unnormalized spin density matrix is explicitly given by the following
expression ($i=q\bar{q},gg,\ldots$):
\begin{eqnarray}
  R_{\alpha\beta,\alpha'\beta'}=
  \sum \<t(k_t,\alpha) \bar{t}(k_{\bar{t}},\beta)X|{\mathcal T}|i\>
  \<t(k_t,\alpha')\bar{t}(k_{\bar{t}},\beta')X|
  {\mathcal T}|i\>^{*},  
\end{eqnarray} 
where the  sum runs over all unobserved degrees of freedom.
The decomposition of the spin density matrix 
with respect to the $t$ and $\bar{t}$ spin spaces
reads:
\begin{equation} 
  R  =  A\ \one\otimes \one 
  + {\bf B}^+\bfsig\otimes\one
  + \one\otimes\bfsig\ {\bf B}^- 
  + C_{ij}\ \sigma^i\otimes\sigma^j \, .
\end{equation}
The polarization of the top quark (antiquark) is encoded in ${\bf B}^\pm$,
e.g.
\begin{equation}
 {\bf P}_{t}  \equiv 2\langle{\bf S}_{t}\rangle={{\bf B}^{+}\over A}
  ={{\rm Tr}\left[R\bfsig\otimes \one\right]\over {\rm Tr}\left[R\right]} ,
\end{equation} 
where ${\bf S}_t$ denotes the top quark spin operator. 
The matrix $C$ encodes the spin correlations 
of the top quark and antiquark:
\begin{equation}
 4\langle{ S}_{t,i}
  { S}_{\bar{t},j}\rangle=
  {{\rm Tr}\left[R\sigma_i\otimes \sigma_j\right]
  \over {\rm Tr}\left[R\right]} = {C_{ij}\over A}\, .
\end{equation}
\subsection{Normal polarization}
If only strong interactions are taken
into account then the polarization of $t$ and $\bar{t}$ 
in $pp,p\bar{p}\to t\bar{t}X$ can only be normal
to the event plane due to parity invariance of QCD.
Normal polarization requires absorptive parts in the scattering
amplitude, i.e. one-loop diagrams with discontinuities.
For the two initial states $i=q\bar{q},gg$,
\begin{equation}
  {\bf P}_{t}^i = {\bf P}_{\bar{t}}^i = b_3^i(y,\hat{s})\hat{\bf{n}},
\end{equation}
where $\hat{\bf{n}}$ is the unit vector normal to the event plane and
$y=\cos\theta$ with $\theta$ denoting the scattering angle of the top quark
in the parton center-of-mass frame.
The normal polarization at the parton level is a percent effect
\cite{Bernreuther:1996cx,Dharmaratna:1990jr}. Several observables
to study this effect
were proposed and studied in ref. \cite{Bernreuther:1996cx}.
At the Tevatron, and probably even at the LHC, the normal polarization 
of the top quark induced by QCD
will be very difficult to observe. 
This means that it provides a sensitive probe of non-standard strong
rescattering effects in hadronic top quark pair production.
\subsection{Spin correlations at leading order}
The correlation between two observables ${{O_1},\ {O_2}}$ is defined as
\begin{equation}
  {\rm corr}({O_1},{O_2})=
  \frac{\langle{O_1}{O_2} \rangle-
    \langle{O_1}\rangle\langle{O_2}\rangle}{\delta{O_1} 
\delta{O_2}},
\end{equation}
with $\delta{O_i}=\sqrt{\langle{O_i^2}\rangle- \langle{O_i}\rangle^2}$.
Assuming parity invariance
we have for hadronic production of top quark pairs at leading order 
(i.e., no absorptive parts):
\begin{equation}
  {\langle{{S}_t^i}\rangle}=
  {\langle{{S}_{\bar{t}}^j}\rangle=0}
\end{equation}
and therefore
\begin{equation}
  {\rm corr}({{S}_t^i},{{S}_{\bar{t}}^j})
  =4\langle{{S}_t^i} 
  {{S}_{\bar{t}}^j}\rangle={C_{ij}\over A}.
\end{equation}
For the process $q\bar{q}\to t\bar{t}$, the spin correlation matrix
at LO is quite simple:
\begin{eqnarray}
  \frac{C_{ij}^{q\bar{q}}}{A^{q\bar{q}}}=\frac{1}{3}\delta_{ij}
  +\frac{2}{2-\beta^2(1-y^2)}\left[\left({\hat{d}_i\hat{d}_j}-
      \frac{1}{3}\delta_{ij}\right)\right.
  + \left.
    \beta^2(1-y^2)\left(\hat{d}_i^{\perp}\hat{d}_j^{\perp}-
      \frac{1}{3}\delta_{ij}\right)
  \right]
\end{eqnarray}
with
\begin{equation}
  {{\hat{\bf d}}}={\frac{1}{\sqrt{1+(\gamma^2-1)y^2}}
    \left[\gamma y \hat{\bf k}_t+\sqrt{1-y^2}\hat{\bf k}_t^{\perp}\right]},
\end{equation}
where $\beta=(1-4m_t^2/s)^{1/2},\ \gamma=(1-\beta^2)^{-1/2}$, and
$\hat{\bf k}_t$ denotes the direction of the top quark. Further, 
${\bf k}^{\perp}=\hat{\bf p}_q-y\hat{\bf k}$ and 
${\bf d}^{\perp}=\hat{\bf p}_q-(\hat{\bf p}_q\cdot \hat{\bf d})\hat{\bf d}$,
where $\hat{\bf p}_q$ denotes the quark direction.

For $q{\bar q} \to t{\bar t}$ 
the direction ${{\hat{\bf d}}}$ is the optimal spin basis at leading
order
QCD, because
at LO:
\begin{equation}
  {{\frac{\hat{d}_iC_{ij}^{q\bar{q}}\hat{d}_j}
      {A^{q\bar{q}}}=1}}.
\end{equation} 
For any ${\beta}$
and ${y}$, the ${{t\bar{t}}}$ spins are 
100\% correlated
w.r.t. to this basis \cite{Mahlon:1997uc}. Following the nomenclature
of  ref. \cite{Mahlon:1997uc} we shall call this basis the off-diagonal basis
in the following. For the Tevatron, where this choice of spin axis is useful
there is an equally efficient but simpler possibility.   
At threshold, the top quark pair is in a $^3S_1$  state and we have
\begin{equation}
  {\hat{\bf d}} {\buildrel {\beta \to 0} \over \longrightarrow}
  \hat{\bf p}_q.
\end{equation} 
This suggests that
at the Tevatron the direction  $\hat{\bf p}$
of, say, the proton beam is an equally  good choice of spin axis. 

At high energies, helicity is conserved and we have
\begin{equation}
  \hat{\bf d} {\buildrel {\beta \to 1} \over \longrightarrow} 
  \hat{\bf k}_t.
\end{equation}
\par
For the process ${gg\to t\bar{t}}$, no optimal spin basis exists.
The LO expression for the spin correlation matrix is quite lengthy, 
and we therefore discuss only the limiting cases. At threshold,  
the top quark pair is  in a $^1S_0$ state and
\begin{equation}
  {\frac{C^{gg}_{ij}}{A^{gg}}} 
  {\buildrel {\beta \to 0} \over \longrightarrow} 
  -\delta_{ij}.
\end{equation}
At high energies, we have
\begin{eqnarray} 
  {\frac{C^{gg}_{ij}}{A^{gg}}} {\buildrel {\beta \to 1} \over 
    \longrightarrow}
  \frac{1}{3}\delta_{ij}+\frac{2}{1+y^2}\left[
    \left(\hat{k}_{t,i}\hat{k}_{t,j}-\frac{1}{3}\delta_{ij}\right)
  \right. 
  + \left. (1-y^2)
    \left(\hat{k}_{t,i}^{\perp}\hat{k}_{t,j}^{\perp}
      -\frac{1}{3}\delta_{ij}\right)
  \right],
\end{eqnarray}
and helicity conservation is reflected by 
\begin{equation} 
  {\frac{\hat{k}_{t,i}C_{ij}^{gg}
      \hat{k}_{t,j}}{A^{gg}} {\buildrel {\beta \to 1} \over 
      \longrightarrow} 1}.
\end{equation}
\section{Observing spin correlations}
The $t \bar t$ spin correlations show up in 
certain angular distributions/correlations of the top decay
products, e.g. 
for the dileptonic decay channel 
${t\to {\ell^+}\nu b,\ \bar{t}\to {\ell'^-}\bar{\nu}\bar{b}}$
the following distribution is sensitive to the correlations: 
\begin{eqnarray}
  {\frac{1}{\sigma}\frac{d^2\sigma(h_1h_2\to
        t{\bar t}X\to \ell^+\ell'^-X)}
    {d\cos \theta_+ \cos \theta_-}
    =\frac{1}{4} ( 1
    +{B_1}\cos\theta_++{B_2}\cos\theta_-
    -C \cos\theta_+\cos\theta_- )}. \label{doublelepton}
\end{eqnarray}
In eq.~(\ref{doublelepton}), 
${{\theta_+,\theta_-}}$ are the angles of ${{\ell^{\pm}}}$
in the ${{t}}$ (${{\bar{t}}}$) rest frame
with respect to arbitrary  spin quantization axes
${\hat{\bf a},\ \hat{\bf b}}$, e.g.:
\begin{eqnarray}
  \begin{array}[h]{ll}
    {\hat{\bf a}=-\hat{\bf b}=\hat{\bf k}_t} 
    & ({\rm helicity \ basis}), \\ 
    {\hat{\bf a}=\hat{\bf b}=\hat{\bf p}} 
    &({\rm beam \ basis}), \\
  {\hat{\bf a}=\hat{\bf b}=\hat{\bf d}}  &
  ({\mbox{off-diagonal \ basis}}). 
  \end{array}
\end{eqnarray}
The coefficient $C$ reflects the strength of the ${{t\bar{t}}}$
spin correlations for the chosen quantization axes, 
${{-1\le C\le+1}}$ \cite{Kuhn:1983ix}. Further we have 
${{B^{\rm QCD}_1}={B^{\rm QCD}_2}=0}$ if ${\hat{\bf a},\hat{\bf b}}$
are chosen to be in the production plane due to P invariance of QCD.


\section{Spin correlations at NLO}
Within the LPA, the coefficient $C$ of eq.~(\ref{doublelepton}) factorizes:
\begin{equation} 
C=\kappa_+\kappa_-D
\end{equation} 
with the ${{t\bar{t}}}$ double spin asymmetry  
\begin{equation}\label{das}
  D=\frac{\sigma(\uparrow\uparrow)+\sigma(\downarrow\downarrow)
    -\sigma(\uparrow\downarrow)-\sigma(\downarrow\uparrow)}{
    \sigma(\uparrow\uparrow)+\sigma(\downarrow\downarrow)
    +\sigma(\uparrow\downarrow)+\sigma(\downarrow\uparrow)}.
\end{equation}
In eq.~(\ref{das}), ${\sigma(\uparrow\uparrow)}$ denotes the  
hadron cross section 
for top quark pairs
with $t(\bar{t})$ spin parallel to the chosen spin quantization
axis ${\hat{\bf a}(\hat{\bf b})}$ etc.
The numbers 
$\kappa_{\pm}$ are the spin analysing powers 
of charged leptons  in decays 
$t(\bar{t})\to b(\bar{b})\ell^{\pm}\nu(\bar{\nu})$. 
The decay distribution reads 
\begin{equation}
  \frac{1}{\Gamma}\frac{d\Gamma}{d\cos\vartheta_{\pm}}=
  \frac{1}{2}\left(1+\pm\kappa_{\pm}\cos\vartheta_{\pm}\right),\label{decay}
\end{equation} 
where
${{\vartheta_{\pm}}}$   
are the angles of ${{\ell^{\pm}}}$ w.r.t. the ${{t}}$ 
(${{\bar{t}}})$ spin.
\subsection{Spin analysing power of top quark decay products}
If the $t$ or $\bar{t}$ quark decays hadronically, one can use other
decay products as spin analysers. One defines in analogy to eq.~(\ref{decay})
for $t\to bW^+\to b\ell^+\nu$ or $\ bq\bar{q}'$:
\begin{equation}
\frac{1}{\Gamma}\frac{d\Gamma}{d\cos\vartheta}=
\frac{1}{2}\left(1+{\kappa_f}\cos\vartheta\right).
\end{equation}
The leading order results for $\kappa_f$ are given in Table~1.
\begin{table}[ht!]     
\caption{\it Leading order results for the spin analysing power
of top quark decay products. In the last column, `${{q\bar{q}'}}$ jet'
stands for the least energetic non-$b$-quark jet in hadronic $t$ decays
\protect\cite{Czarnecki:1994pu} .}
\begin{center}\renewcommand{\arraystretch}{1.5}
  \begin{tabular}{|c|c|c|c|c|c|}  \hline
    $f$ & ${{\ell^+,\ \bar{d},\ \bar{s}}}$ & 
    ${{\nu,\ u}}$ &
    ${{b}}$ &  
    ${{W^+}}$ &
    {\footnotesize{ ${{q\bar{q}'}}$ jet }}        
    \\
    \hline
    ${{\kappa_f}}$ & 1 & $-$0.31 & $-$0.41 & 0.41 &0.51  \\ 
    \hline  
  \end{tabular}
\end{center}
\end{table}
In order to compute the spin correlation coefficient $C$ in NLO in $\alpha_s$,
we need the QCD corrections to the spin analysing power $\kappa_f$. 
For leptonic decays, the corrections are tiny \cite{Czarnecki:1991pe}
\begin{equation}
  \kappa_+=\kappa_-=1-0.015\alpha_s, 
\end{equation}
implying that the charged lepton is a perfect analyser of top quark spin.
However, since only about  5\%  of the decays of 
${{t\bar{t}}}$ are purely 
leptonic (e,${\mu}$), it is also important to compute the QCD corrections
for hadronic decays ${{t\to bq\bar{q}'}}$. The results are 
\cite{Brandenburg:2002xr}
(using $\alpha_s(m_t)=0.108$):
\begin{eqnarray}
  \kappa_{\bar{d}}&=&1-0.57\alpha_s=0.94, \\
  \kappa_b&=&-0.41\times(1-0.34\alpha_s)=-0.39,\\
  \kappa_j&=&+0.51\times(1-0.65\alpha_s)=+0.47,
\end{eqnarray}
where
${{\kappa_j}}$ is the analysing power of least energetic
non-$b$-quark jet. One observes
that the ${\bar{d},\bar{s}}$-jets
have the highest analysing power, but the reconstruction of their direction
is very difficult due to the low efficiency of charm tagging.
A better choice is to use the ${b}$-jet or the least energetic
non-$b$-quark jet. The NLO results for $C$ in the single lepton channel
are less sensitive due to the  factor  ${{\kappa_{j,b}}}$, but this 
loss in sensitivity is overcompensated
by higher statistics, since one has about  30\% 
$({e+\mu})+X$ single lepton ${{t\bar{t}}}$ decays. 
(Nonleptonic top quark decays at order $\alpha_s$ were also analyzed
in ref.~\cite{Fischer:2001gp}.)

\subsection{Double spin asymmetries at NLO at the parton level}
The spin dependent cross sections that enter the double 
spin asymmetry (\ref{das}) are calculated as a convolution which reads
schematically: 
\begin{equation}
  {\sigma(\uparrow\uparrow) =  {\rm PDF's}\otimes \hat{\sigma}} 
  (\uparrow\uparrow),\ldots 
\label{PDF}
\end{equation}
We renormalize
the top mass in the on-shell scheme, and $\alpha_s$ 
in the  modified minimal subtraction (${\overline{\rm MS}}$) scheme.
Factorization is performed in the ${\overline{\rm MS}}$ scheme,
and we set ${\mu_F=\mu_R=\mu}$.
The results at NLO QCD for the ${\overline{\rm MS}}$ subtracted 
parton cross sections 
${{q\bar{q}\to t\bar{t}(g)}}$,
${{gg\to t\bar{t}(g)}}$, and  
${{q(\bar{q})g\to t\bar{t}q(\bar{q})}}$
with ${{t\bar{t}}}$ spins summed over was computed already more than 10 years
ago \cite{Nason:1988xz,Beenakker:1989bq,Beenakker:1991ma}.
It can be written as follows:
\begin{eqnarray}
  \hat{\sigma}(\hat{s},m_t^2)&=&{{\hat{\sigma}(\uparrow\uparrow)+
      \hat{\sigma}(\downarrow\downarrow)+\hat{\sigma}(\uparrow\downarrow)
      +\hat{\sigma}(\downarrow\uparrow)}} \nonumber \\ 
  &=& \frac{\alpha_s^2}{m_t^2}
  \left\{f^{(0)}(\eta)+4\pi\alpha_s
    \left[{f^{(1)}(\eta)}+{\tilde{f}^{(1)}(\eta)}\ln(\mu^2/m_t^2)
    \right]\right\},
\end{eqnarray}
where $\eta=\hat{s}/(4m_t^2)-1$. In our calculation of the
spin-dependent
cross sections refs.~\cite{Bernreuther:2000yn,Bernreuther:2001bx}
we obtain these
cross sections as  special cases and we
find perfect agreement with the results of 
refs. \cite{Nason:1988xz,Beenakker:1989bq,Beenakker:1991ma}.
An analogous decomposition can be defined for the
numerator of the double spin asymmetry: 
\begin{eqnarray}
  {{\hat{\sigma}\hat{D}}}&=&{{\hat{\sigma}(\uparrow\uparrow)+
      \hat{\sigma}(\downarrow\downarrow)-\hat{\sigma}(\uparrow\downarrow)
      -\hat{\sigma}(\downarrow\uparrow)}} \nonumber \\ 
  &=&
  {\frac{\alpha_s^2}{m_t^2}
    \left\{g^{(0)}(\eta)+4\pi\alpha_s
      \left[{g^{(1)}(\eta)}+{\tilde{g}^{(1)}(\eta)}\ln(\mu^2/m_t^2)
      \right]\right\}}.
\label{sigD}
\end{eqnarray}

The following figures display our results for the functions 
$g^{(0)}, g^{(1)}$ and ${\tilde g}^{(1)}$ calculated for several
spin quantization axes. In eq.~(\ref{sigD}) the coupling  
$\alpha_s$ denotes the six-flavour coupling $\alpha_s^{f=6}$. 
Fig.~1 shows our results for the helicity basis. In Figs. 2 and 3 
 corresponding results for the beam basis and the  
off-diagonal basis are displayed.
\begin{figure}[ht!]
  \unitlength1.0cm
  \begin{center}
    \begin{picture}(6.5,6.5)
      \put(-5.75,0){\psfig{figure=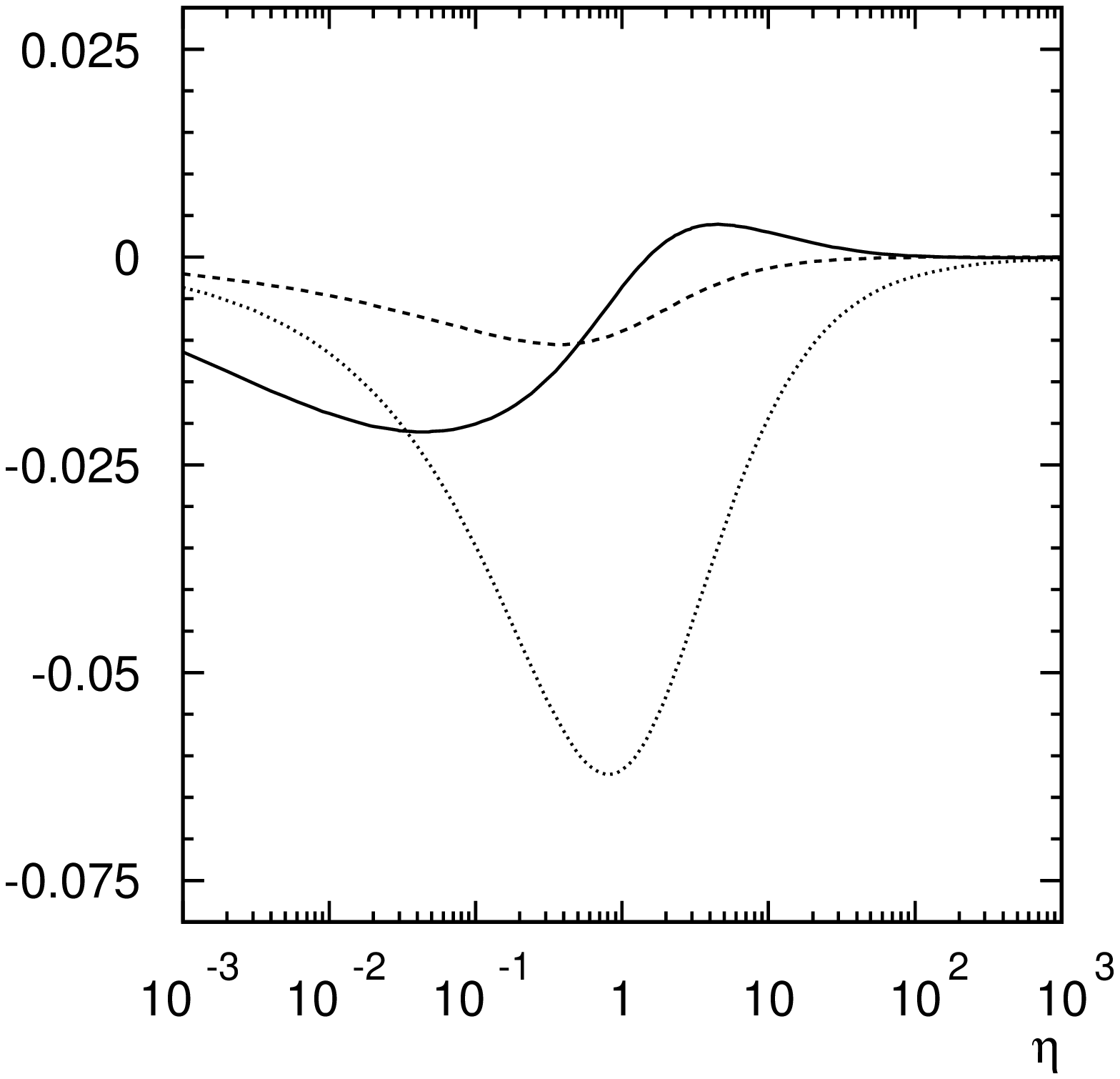,width=6.5cm,height=6.5cm}}
      \put(-0.0,0){\psfig{figure=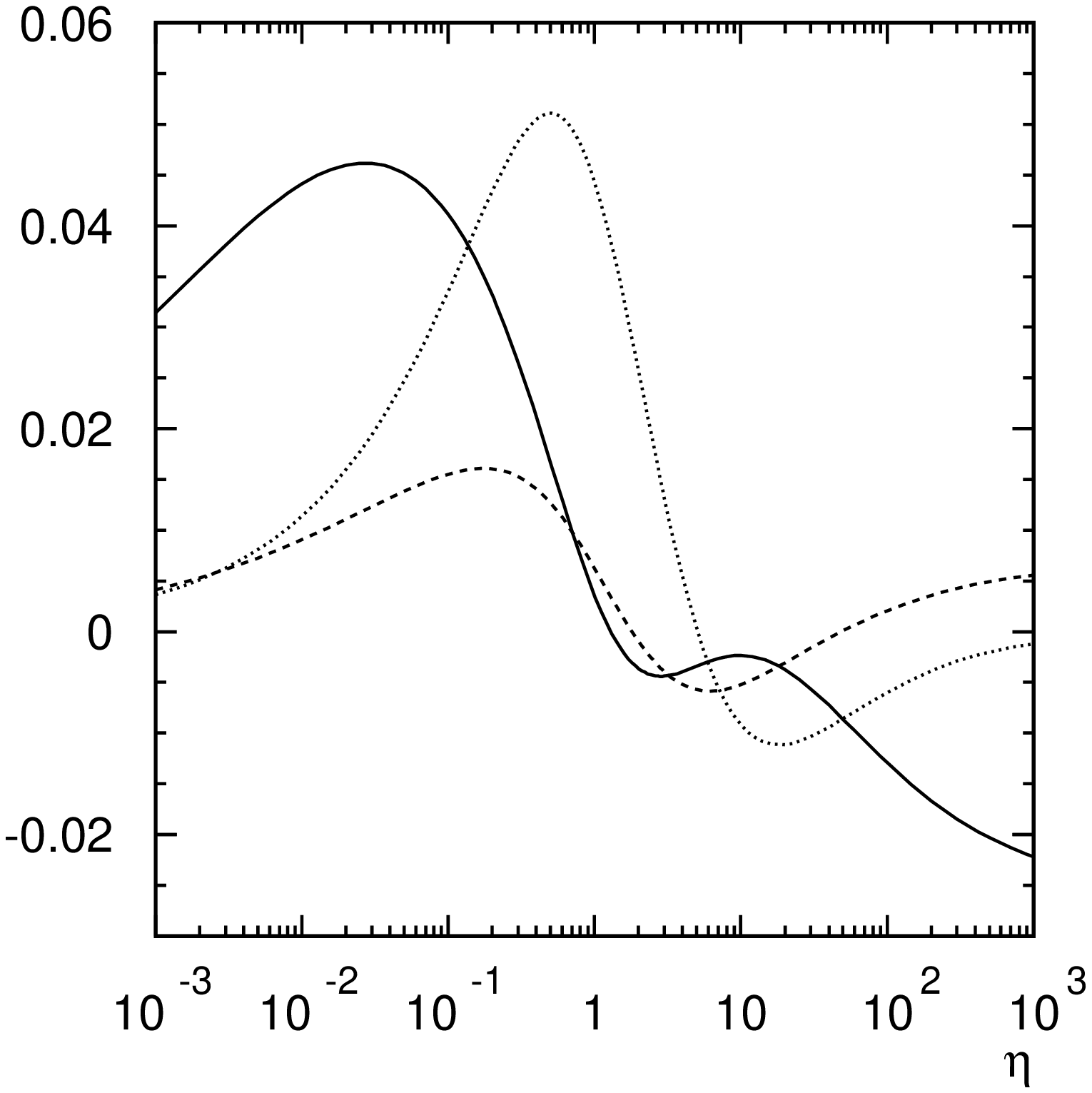,width=6.5cm,height=6.5cm}}
      \put(5.75,0){\psfig{figure=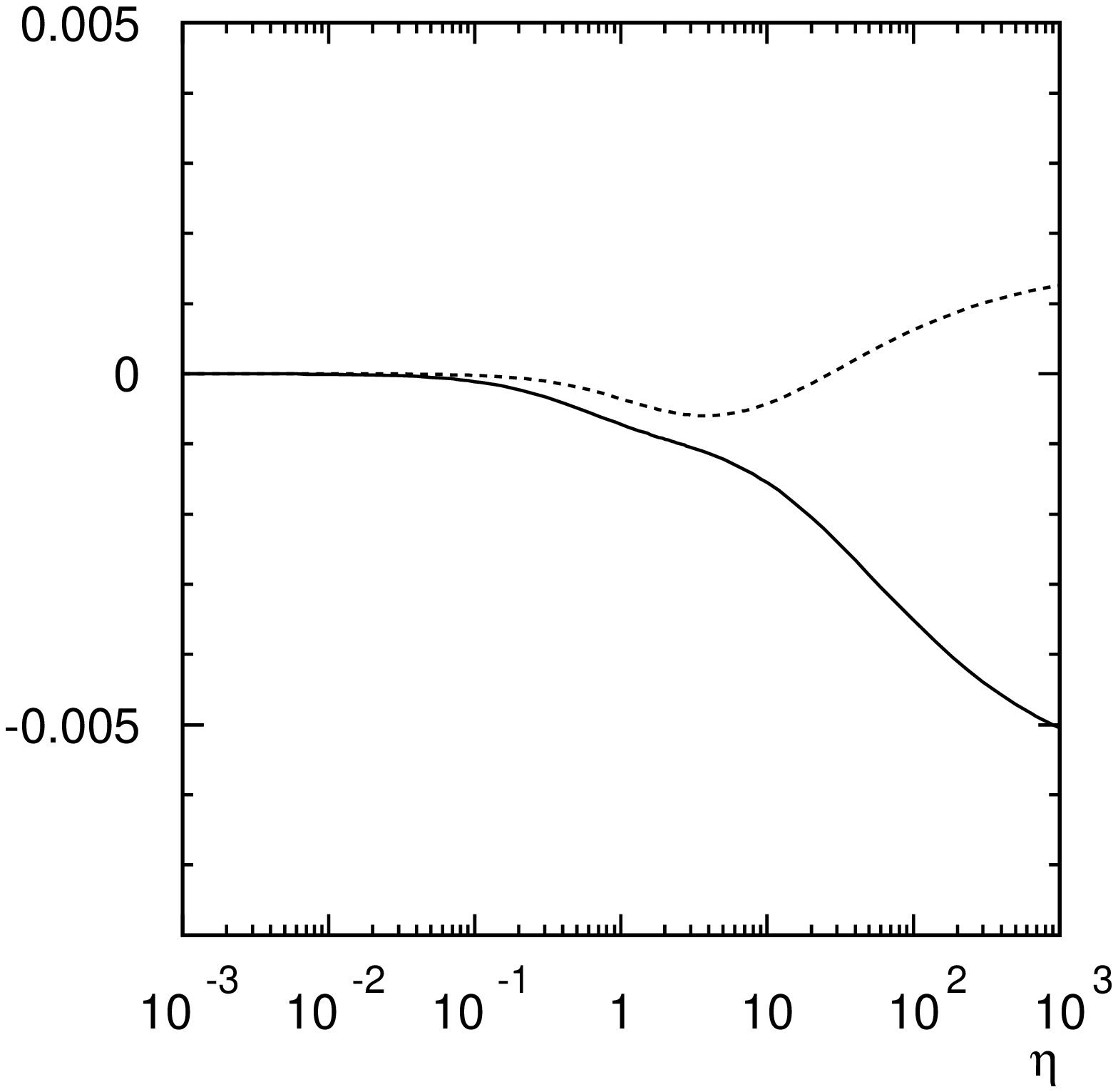,width=6.5cm,height=6.5cm}}
    \end{picture}
    \caption{\it Left: Scaling functions $g^{(0)}(\eta)$ (dotted), 
      $g^{(1)}(\eta)$ (full), and $\tilde{g}^{(1)}(\eta)$ (dashed) 
      in the helicity basis
      for the process $q\bar{q}\to t\bar{t}(g)$. Middle: The same 
      for the process $gg\to t\bar{t}(g)$.
      Right: The functions $g^{(1)}(\eta)$ (full), 
      and $\tilde{g}^{(1)}(\eta)$ (dashed)  for the 
      process  $qg\to qt\bar{t}$ \cite{Bernreuther:2000yn,Bernreuther:2001bx}.}
    \label{fig:o2}
  \end{center}
\end{figure}

\begin{figure}[ht!]
  \unitlength1.0cm
  \begin{center}
    \begin{picture}(6.5,6.5)
      \put(-5.75,0){\psfig{figure=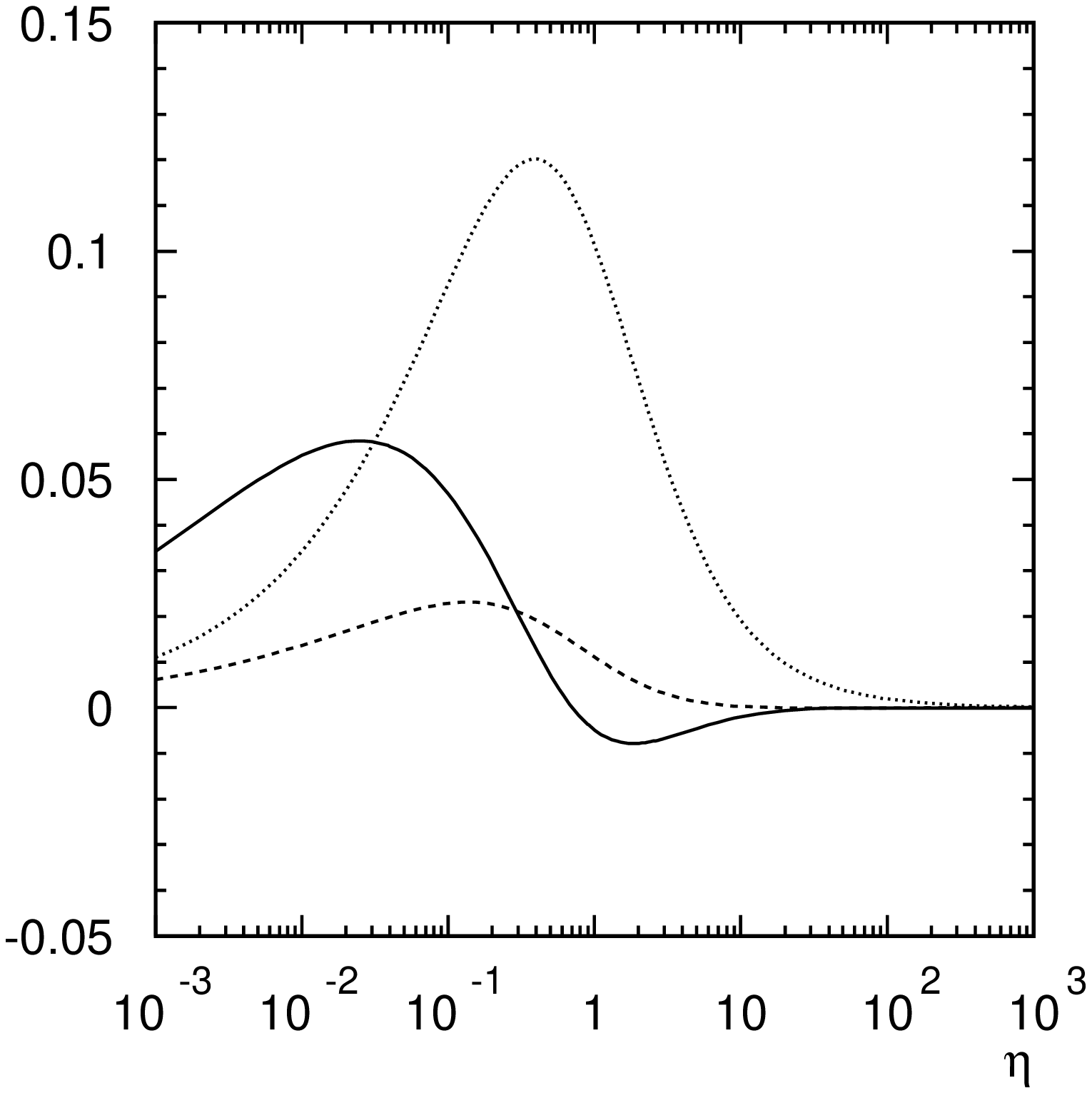,width=6.5cm,height=6.5cm}}
      \put(-0.0,0){\psfig{figure=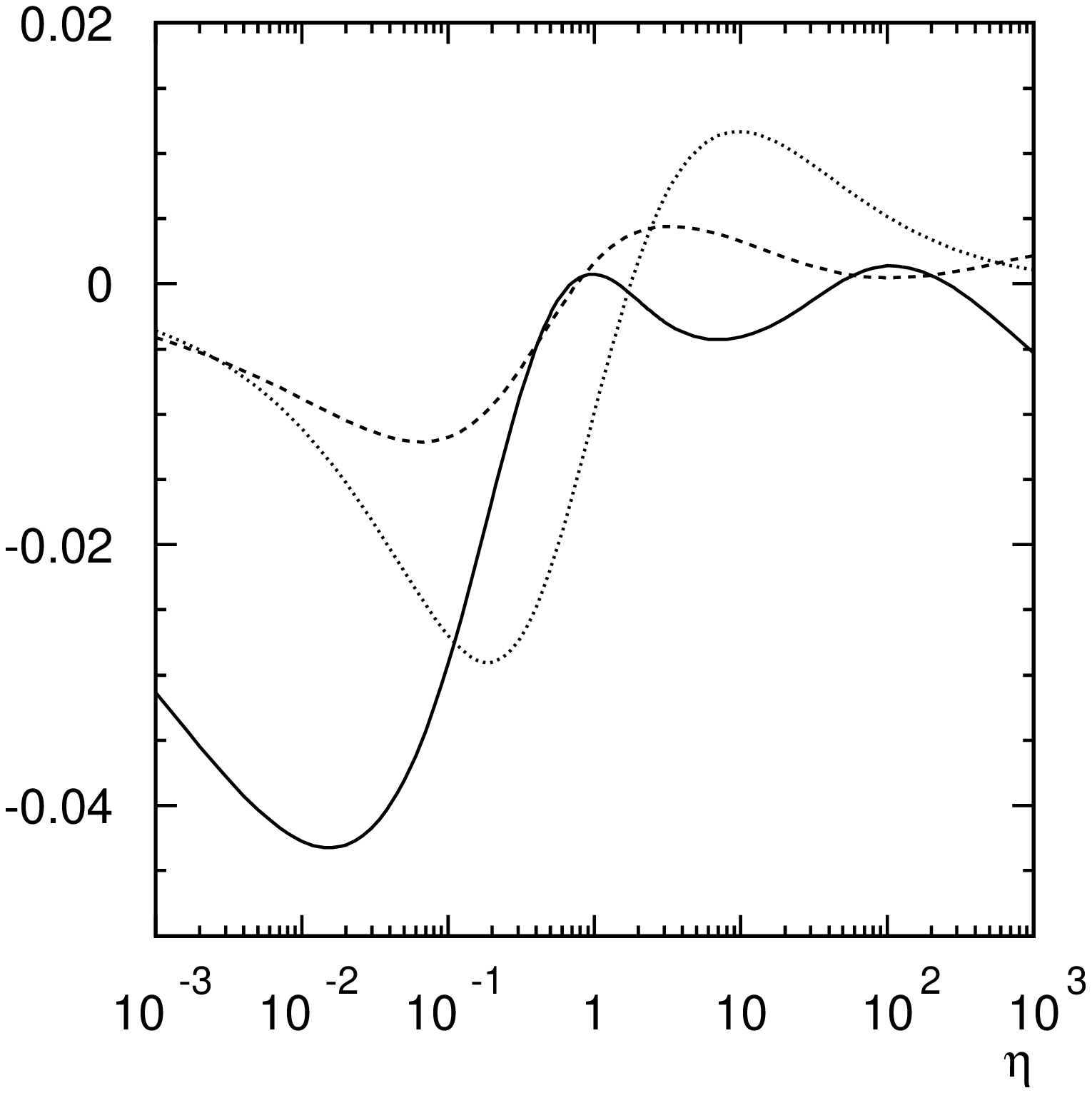,width=6.5cm,height=6.5cm}}
      \put(5.75,0){\psfig{figure=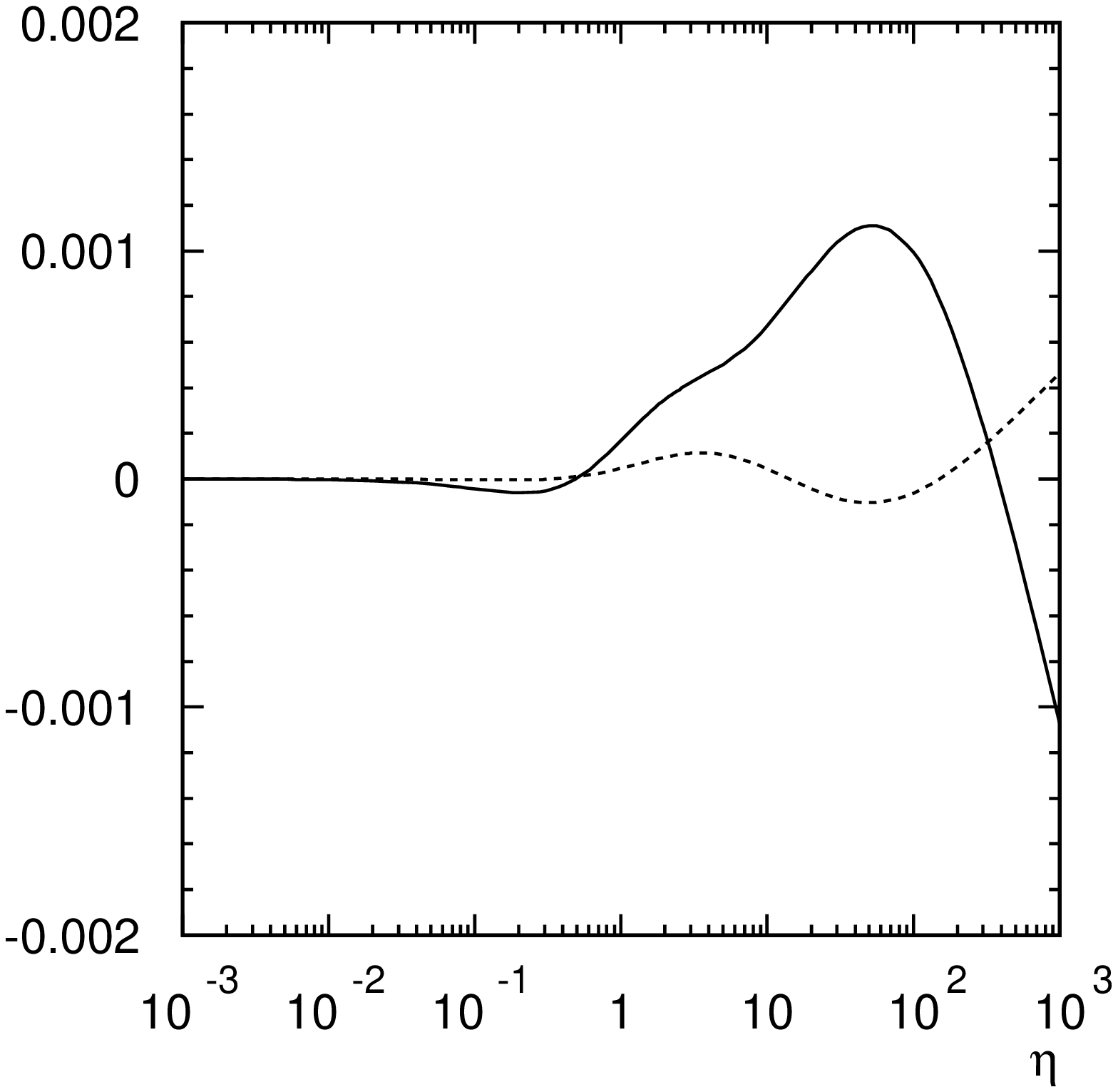,width=6.5cm,height=6.5cm}}
    \end{picture}
    \caption{\it The same as Fig.~1, but for the beam basis 
    \cite{Bernreuther:2000yn,Bernreuther:2001bx}.}
    \label{fig:o5}
  \end{center}
\end{figure}

\begin{figure}[ht!]
  \unitlength1.0cm
  \begin{center}
    \begin{picture}(6.5,6.5)
      \put(-5.75,0){\psfig{figure=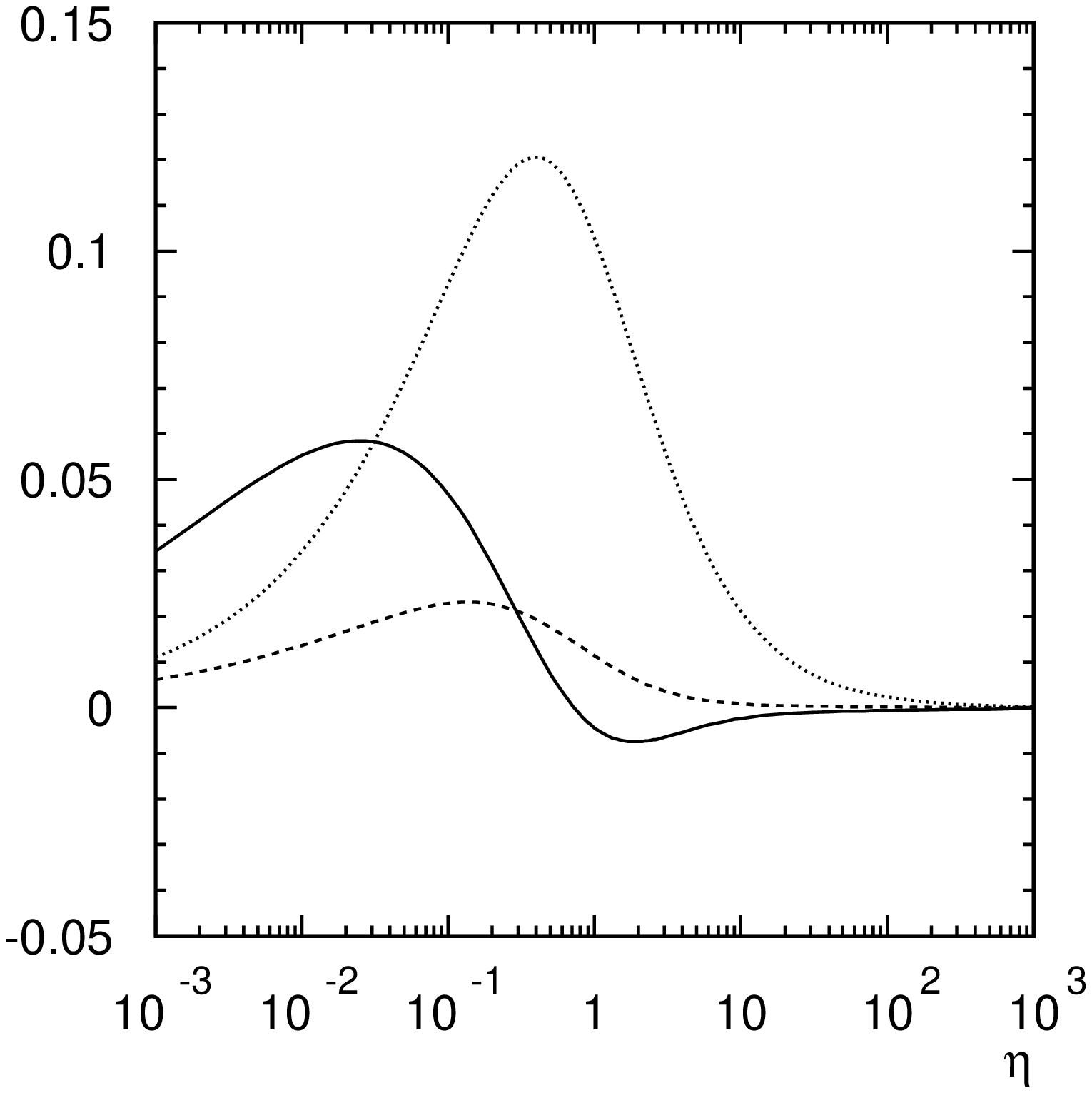,width=6.5cm,height=6.5cm}}
      \put(-0.0,0){\psfig{figure=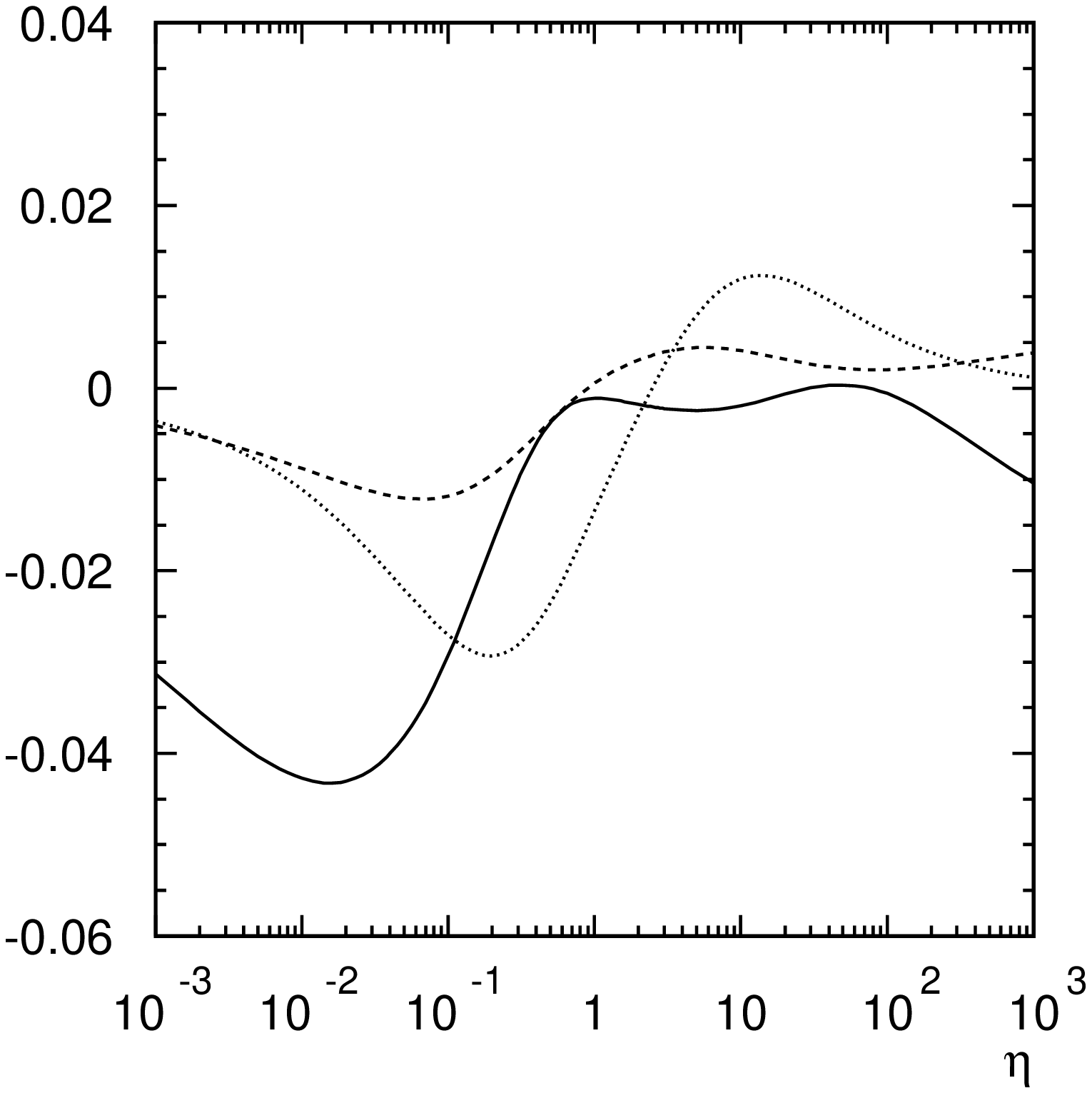,width=6.5cm,height=6.5cm}}
      \put(5.75,0){\psfig{figure=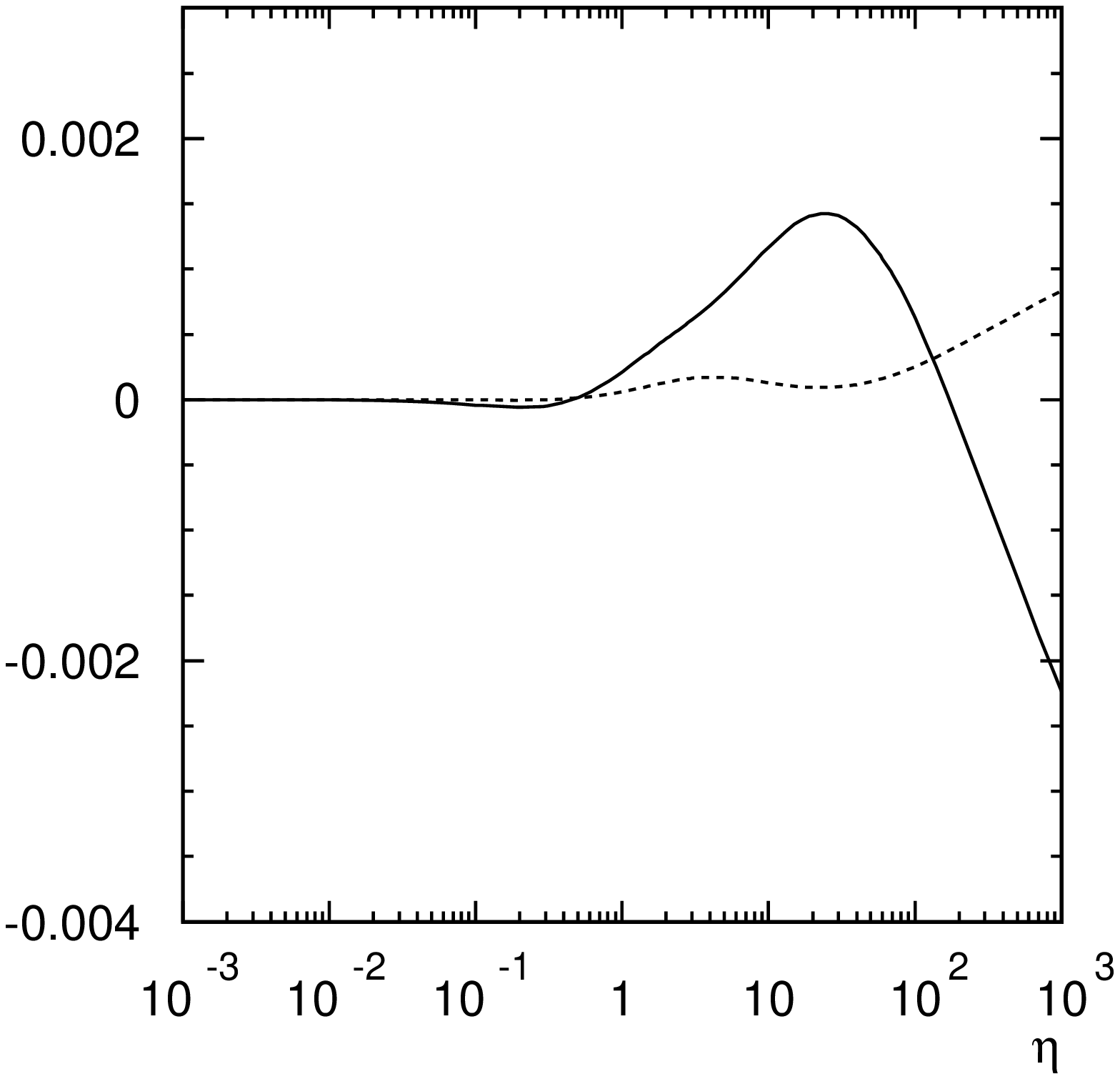,width=6.5cm,height=6.5cm}}
    \end{picture}
    \caption{\it The same as Fig.~1, but for the optimal basis
\cite{Bernreuther:2000yn,Bernreuther:2001bx}.}
    \label{fig:o6}
  \end{center}
\end{figure}

It is natural to express  the above 
partonic
cross sections in terms  of the ${\overline{\rm MS}}$ coupling
$\alpha_s^{f=6}$ in $f=6$ flavour QCD. However, for the evaluation
of hadronic observables, e.g. eq.~(\ref{PDF}), 
the parameter change  $\alpha_s^{f=6} \to \alpha_s^{f=5}$ using 
the standard  ${\overline{\rm MS}}$ relation 

\begin{equation}
  \alpha_s^{f=6} (\mu_R) = \alpha_s^{f=5} (\mu_R)
\left[1 - \frac{1}{3\pi}\alpha_s^{f=5} (\mu_R) \ln\left({m_t\over \mu_R}\right)
+ {\cal O}(\alpha_s^2)\right]
\end{equation} 
is necessary, in order to
make contact with the physics and formalism incorporated in  the PDF libraries.
The evolution of $\alpha_s^{f=5}$
is determined  by the  beta function with 
$n_{f}^{\rm light}=5$ flavours.

\subsection{NLO results for differential decay distributions}  
We now turn to the coefficient $C$ in the 
double angular distribution (\ref{doublelepton}).
For definiteness we discuss the spin correlation $C$ only for the
dilepton channels. These results were obtained in ref. \cite{Bernreuther:2001rq}.
The numbers in Table~2 are obtained
for $\mu_F=\mu_R=m_t=$175 GeV and using the PDF's 
CTEQ5L (LO) and  CTEQ5M (NLO) \cite{Lai:1999wy}.
\begin{table}[h]
  \caption{\it Coefficient $C$
    of the double distribution (\ref{doublelepton})
    to leading and
    next-to-leading order  in $\alpha_s$ for the helicity basis, the
    beam basis (where the proton beam is taken as the spin quantization axis)
    and the off-diagonal basis. 
    The parton distribution
    functions of ref.~\protect\cite{Lai:1999wy} were used 
    choosing the renormalization scale
    $\mu_R$ equal to the factorization scale $\mu_F$ =
    $m_t$ = 175 GeV.}
  \begin{center}\renewcommand{\arraystretch}{1.5}
    \begin{tabular}{|c|cc|cc|} \hline
      &\multicolumn{2}{c|}{$p\bar p$ at $\sqrt{s}=2$~TeV }
      &\multicolumn{2}{c|}{$pp$ at $\sqrt{s}=14$~TeV }
      \\
      \cline{2-5}
      & LO & NLO & LO & NLO\\  \hline \hline
      ${\rm C}_{\rm hel.}$ & $-0.456$& $-0.389$ & $\hphantom{-}0.305$  & 
      $\hphantom{-}0.311$\\
      ${\rm C}_{\rm beam}$ & $\hphantom{-}0.910$&  $\hphantom{-}0.806$ & 
      $-0.005$ & $-0.072$\\
      ${\rm C}_{\rm off.}$ & $\hphantom{-}0.918$ & $\hphantom{-}0.813$ & 
      $-0.027$ & $-0.089$\\ \hline
    \end{tabular}
    \vspace*{1em}
    \label{tab:cteq5}
  \end{center}
\end{table}
At the Tevatron, the 
dilepton spin correlations are large in the beam and the 
off-diagonal basis. Thus one sees there is practically
no difference between these two choices as far as the sensitivity
to QCD-induced spin correlations is concerned; yet the beam basis may
be simpler to implement in the analysis of experimental data.
The QCD corrections are about  $-10$\%. At the
LHC, the  beam and off-diagonal bases are bad choices (due to the dominance
of $gg\to t\bar{t}$). Here the helicity basis is a  good choice, and 
the QCD corrections are small. 
The inclusion of the QCD corrections
reduces the dependence of the $t\bar{t}$ cross section on
the renormalization and factorization scales significantly.
The same is true
for  the product $\sigma{C}$, as can be seen from Fig.~4. 
\begin{figure}[ht!]
  \unitlength1.0cm
  \begin{center}
    \psfig{figure=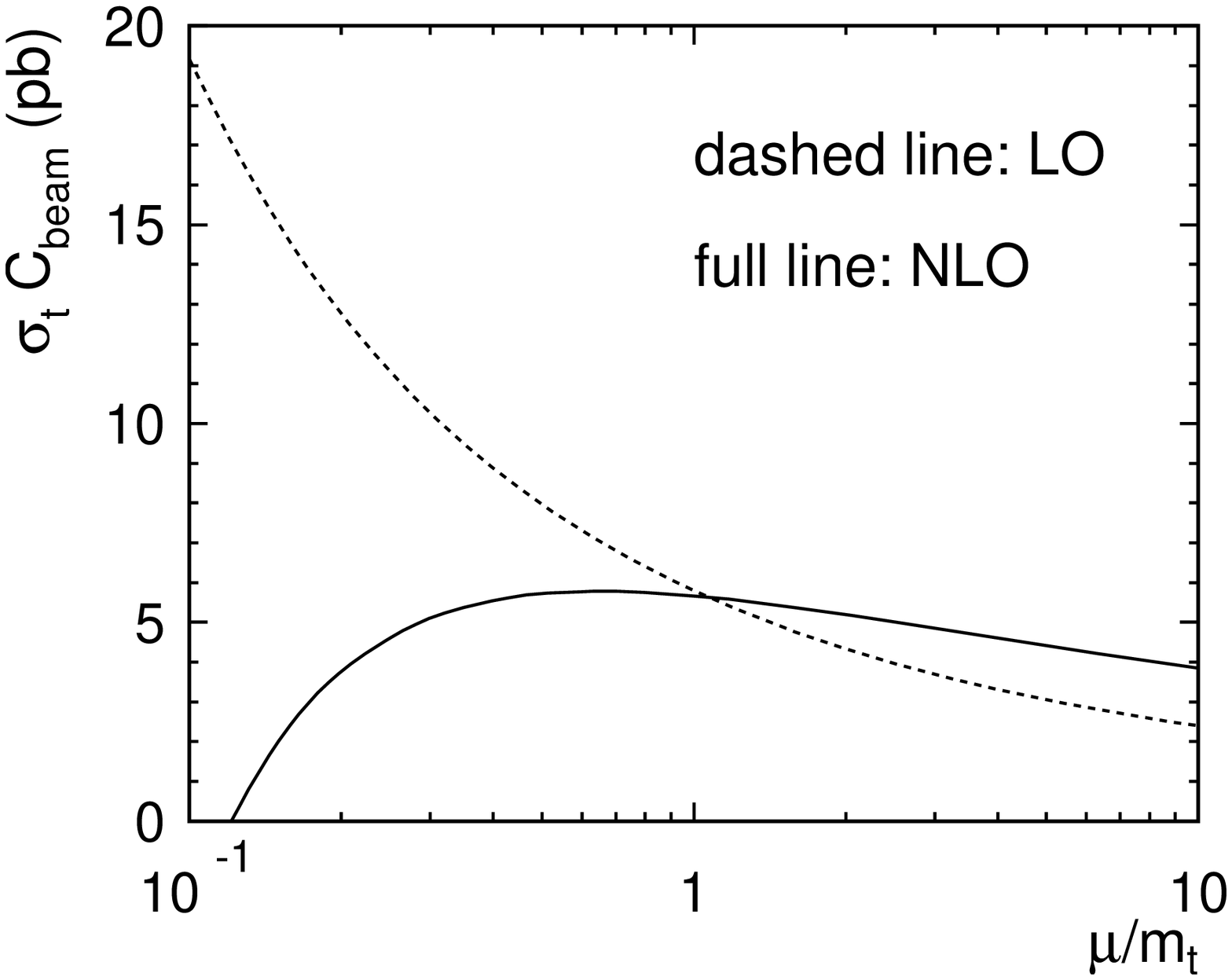,width=0.45\textwidth}\hfill
    \psfig{figure=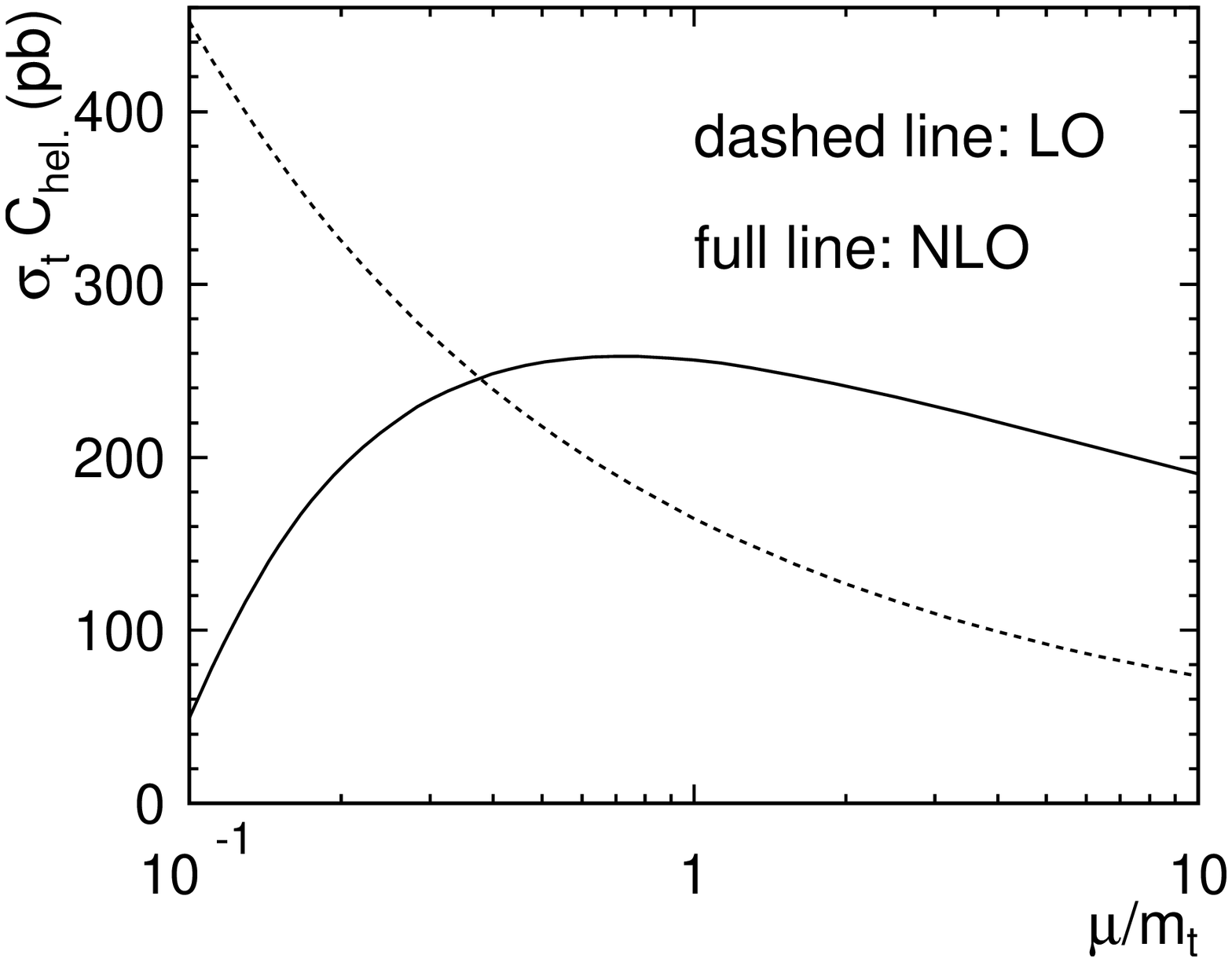,width=0.45\textwidth}
    \caption{\it  Left: Dependence of
      $\sigma{C}_{\rm beam}$ at LO (dashed line) and  at NLO
      (solid line) on $\mu=\mu_R=\mu_F$
      for $p \bar p$ collisions at $\sqrt{s}=$ 2 TeV,
      with PDF  of  ref.~\protect\cite{Lai:1999wy}. Right:  Same, but for
      $\sigma{C}_{\rm hel.}$ 
      for $pp$ collisions at $\sqrt{s}=$ 14 TeV.}
    \label{fig:mu}
  \end{center}
\end{figure} 
\begin{table}[ht!]
\vspace{-0.3cm}
\caption{\it Upper part: Dependence on the scale 
$\mu$ of the correlation coefficients
  $C$, computed at NLO with the PDF  of ref.\protect\cite{Lai:1999wy}. 
  Lower part:
  Correlation coefficients $C$ 
  at NLO for $\mu=m_t$ and different sets
  of parton distribution functions:  GRV98\protect\cite{Gluck:1998xa}, 
  CTEQ5\protect\cite{Lai:1999wy}, 
  and MRST98 (c-g)\protect\cite{Martin:1998sq}.} 
\par
\begin{center}
  \renewcommand{\arraystretch}{1.5}
  {\begin{tabular}{|ccccc|} \hline
      & \multicolumn{3}{c}{Tevatron}
      & LHC \\
      $\mu$   & $C_{\rm hel.}$ &  $C_{\rm beam}$ &
      $C_{\rm off.}$ &  $C_{\rm hel.}$ \\ \hline
      $m_t/2$ & $-0.364$   & 0.774    & 0.779  &   0.278
      \\
      $m_t$   & $-0.389$   & 0.806  & 0.813 &  0.311
      \\
      $2m_t$ & $-0.407$  & 0.829 & 0.836  &  0.331
      \\ \hline \hline
      PDF   & $C_{\rm hel.}$ &  $C_{\rm beam}$ &
      $C_{\rm off.}$ &  $C_{\rm hel.}$ \\ \hline
      GRV98 & $-0.325$   & 0.734    & 0.739  &  0.332
      \\
      CTEQ5   & $-0.389$   & 0.806  & 0.813 &  0.311
      \\
      MRST98 & $-0.417$  & 0.838 & 0.846  &  0.315
      \\ \hline
    \end{tabular}}\label{tab:mudep}
\end{center}
\end{table}
Table~3 shows the dependence of the NLO results for $C$ on the scale
$\mu$ (upper part) and on the choice of the PDFs (lower part).
At the Tevatron the spread of results for different PDFs is larger
than the scale uncertainty: The results for $C$ using 
the CTEQ5 and MRST98  distributions 
agree up to a few percent, 
but the difference between GRV98 and MRST98 at the Tevatron is about  
$10\%$.
The main reason for this strong dependence on the PDFs is that the 
contributions from the $gg$  and the $q\bar{q}$ initial state enter
with a different sign. 
This offers the interesting possibility to constrain the  PDFs
by measuring $t\bar{t}$ spin correlations.

In all results above we used $m_t=175$ GeV. A variation of 
$m_t$ from $170$ to $180$ GeV changes the results for the 
Tevatron, again for $\mu=m_t$ and PDFs of ref.~\cite{Lai:1999wy} as follows: 
$C_{\rm hel.}$ varies from $-0.378$ to $-0.397$, $C_{\rm beam}$
from $0.790$ to $0.817$, and  $ C_{\rm off.}$ from $0.797$ to
$0.822$. For the LHC,  $C_{\rm hel.}$ changes by less than a percent.  

The results above have been obtained without imposing any kinematic cuts.
Standard cuts on the top quark transverse momentum and rapidity only have
a small effect on $C$: For the Tevatron, demanding 
$|{\bf k}_{t,\bar{t}}^T|>15$ GeV and $|r_{t,\bar{t}}|<2$ leads to 
the following results: 
$C_{\rm hel.}=-0.386,\ C_{\rm beam}=0.815,\ C_{\rm off.}=0.823$.
For the LHC, when imposing the cuts $|{\bf k}_{t,\bar{t}}^T|>20$ GeV 
and $|r_{t,\bar{t}}|<3$, we find $C_{\rm hel.}=0.295$.
\par
A tool for simulating these spin correlations at leading order in the
QCD coupling exists \cite{Slabospitsky:2002ag}.
Apart from the double differential distribution (\ref{doublelepton})
there are also other  distributions  which are  sensitive 
to spin correlations \cite{BBSI}. Finally
let us mention that  
there has been a first 
measurement of spin correlations in the off-diagonal 
 basis by the D0 collaboration 
\cite{Abbott:2000dt}. It is 
based on six dilepton events from Run I. They find
\begin{equation}
  C_{\rm off.}>-0.25 \ \  @ \ \ 68\% \ {\rm confidence \ level}
\end{equation}
This demonstrates that top quark spin correlations can be studied
already at the Tevatron.
It is expected that the correlations can be established at the $2\sigma$ 
level using Run II data.
\section{Conclusions}
QCD-induced 
spin correlations of top quark  pairs  produced at hadron colliders  are 
large effects. They can be studied at the Tevatron and LHC. 
For the Tevatron the QCD corrections to the leading order predictions 
are sizeable but under
control. The degree of correlation depends on the choice of the spin
quantization axis.
We have shown that using, at the Tevatron, 
 the direction of one of the hadron beams
as spin quantization axis is as efficient as the off-diagonal axis which has
received much attention in the literature.
Spin correlations are suited to study in detail the interactions of 
top quarks. As we have pointed out 
they should, in particular,
 be a useful tool for  constraining  PDFs. 
Taking the PDFs, once they have been determined with sufficient precision,
as input, spin correlations will be an important  tool 
for the search for new effects in top quark pair production.
Future work on the theory side will include the 
implementation of the NLO matrix elements in an event generator,
a study of non-factorizable corrections, and 
the resummation of Sudakov-type large logarithms for the spin-dependent
cross sections.
\section*{Acknowledgments}
A.B. would like to thank the organizers of the Epiphany conference
for their hospitality. 


\end{document}